\documentclass[aps,prd,amsmath,amssymb,12pt]{revtex4}
\usepackage{graphicx}
\usepackage{bm}
\def\journal#1#2#3#4{{#1} {\bf #2}, #3 (#4)}
\newcommand{\be}{\begin{equation}}
\newcommand{\ee}{\end{equation}}
\newcommand{\bea}{\begin{eqnarray}}
\newcommand{\eea}{\end{eqnarray}}
\newcommand{\hf}{\frac12}
\newcommand{\nn}{\nonumber\\}
\def\eq#1{(\ref{#1})}
\def\la{\langle}
\def\ra{\rangle}

\def\Tr{{\mathrm{Tr}}}
\def\ord#1{{\cal O}\left(#1\right)}

\def\v#1{{\bm{#1}}}

\def\hx{\hat x}
\def\hy{\hat y}

\def\ih{\frac{i}{\hbar}}

\begin{document}
\title{A Gaussian density matrix under decoherence and friction}
\author{Janos Polonyi}
\email{polonyi@iphc.cnrs.fr}
\affiliation{Strasbourg University, CNRS-IPHC, 23 rue du Loess, BP28 67037 Strasbourg Cedex 2, France}

\begin{abstract}
The time evolution of a Gaussian density matrix of a one dimensional particle, generated by a quadratic, $\ord{\partial_t^2}$ effective Lagrangian, describing a harmonic potential, a friction force and decoherence, is studied within the Closed Time Path formalism. The density matrix converges to an asymptotic form, given by a completely decohered thermal state with an $\ord{\hbar}$ temperature in the translation invariant case. The time evolution of the state of a harmonic oscillator is followed numerically. The asymptotic density matrix, the fixed point of the master equation, is found analytically and its dependence on the oscillator frequency, the friction constant and the decoherence strength is explored.
\end{abstract}
\maketitle

\section{Introduction}
Open systems represent a specially difficult problem in Quantum Mechanics owing to the entanglement with their environment. The main technical difficulty consists of the systematic derivation of dissipative phenomenas, such as the friction force and the decoherence, arising in realistic open systems. The simplest model where these features appear is the quantum Brownian motion, a text particle moving in a gas.

One way to approach this problem is to rely on the quantum kinetic theory where the time evolution of the density matrix of the test particle can be approximated by a master equation. For the present context such an equation has first been derived to display decoherence \cite{zeh} and the inclusion of dissipation followed later \cite{diosi,adler}. The systematic derivation of the master equation, using the collision cross section of the test particle has been developed, as well \cite{vacchini,lanz,vacchinie,hornbergerk}. The traditional many-body formalism was used in ref. \cite{dodd} to arrive at a master equation.

Another approach to the Brownian motion is the model building. The simplest, harmonic model consists of infinitely many, linearly coupled harmonic oscillators \cite{agarwal,calderialeggett,unruh} and the appropriate formalism to find the effective theory for the system coordinate is the Closed Time Path (CTP) scheme, introduced in quantum field theory some time ago \cite{schw} and generalized to different area of Condensed Matter Physics \cite{keldysh,rammer,kamenev} and Particle Physics \cite{calzetta}. The dissipative harmonic model \cite{dekker} has been studied in this formalism \cite{grabert} and the master equation for the reduced density matrix of the system with memory term has been derived \cite{hu9293}. The Markovian approximation, satisfying the physical requirements of the density matrix has been worked out for high enough temperature in refs. \cite{munro,breuerp,diosip}.

A third way to describe the dynamics of the test particle, the direct calculation of the effective action \cite{guinea} in the CTP formalism \cite{hedegard}, has been advanced  by finding the effective Lagrangian in the leading order of the Landau-Ginzburg double expansion where the role of the small parameters is played by the amplitude and the frequency of the distortion of the particle trajectories by the gas \cite{gas}. One finds in this manner a bridge between the microscopic description and the model building. 

We report in this paper the detailed study of the effective theory \cite{gas} in an exactly solvable case, by assuming that the test particle moves either freely in the gas or in the presence of an external harmonic potential. The path integral expression of the reduced density matrix is evaluated for a Gaussian wave packet. The time evolution of the Gaussian density matrix has already been studied by integrating the equation of motion for the density matrix \cite{joos,morikawa}. The procedure, followed here is based on a more general effective Lagrangian which preserves the positivity of the density matrix.

The paper starts with a short recall of the relevant features of the CTP formalism in section \ref{ctps}, followed in section \ref{effths} by the presentation of the effective Lagrangian of the test particle in a gas, derived in the one-loop approximation and in the leading order Landau-Ginzburg expansion. The path integral representation of the Liouville propagator for the density matrix is given in section \ref{propls}. Section \ref{waveps} contains our mains results, namely the trajectory of a Gaussian density matrix for a free particle or a harmonic oscillator with friction force and decoherence. The discussion of the main features of the asymptotic, relaxed states is given in section \ref{asymptss}. Our conclusions are listed in section \ref{concls}.

\section{CTP formalism}\label{ctps}
A fundamental quantity in the study of the closed dynamics of a particle is the transition amplitude between the pure states $|\psi_i\ra$ and $|\psi_f\ra$, given at times $t_i$ and $t_f$, respectively,
\be\label{trampl}
\la\psi_f|U(t_f,t_i)|\psi_i\ra=\int D[x]\la\psi_f|x(t_f)\ra e^{\ih S[x]}\la x(t_i)|\psi_i\ra,
\ee
where $U(t_f,t_i)$ denotes the time evolution operator. The functional integration on the right hand side is over trajectories $x(t)$ with $t_i<t<t_f$. In the case of a mixed state the density matrix at time $t_f$ is given by
\be\label{densm}
\rho(x^+_f,x^-_f)=\int D[\hx]\rho_i(x^+(t_i),x^-(t_i))e^{\ih S[\hx]},
\ee
where the integration is over pairs of trajectories, $\hx(t)=(x^+(t),x^-(t))$, $t_i<t<t_f$, with fixed end points, $x_f^\pm=x^\pm(t_f)$, and the action,
\be\label{ctpaction}
S[\hx]=S[x^+]-S[x^-]+S_\epsilon[\hx],
\ee
contains an infinitesimal imaginary term,
\be\label{regeps}
S[\hx]=i\frac\epsilon2\int_{t_i}^{t_f}dt[x^{+2}(x)+x^{-2}(x)],
\ee
to assure the convergence for large $x$. 

The distinguished feature of this formalism is the reduplication of the degrees of freedom, $x\to x^\pm$, reflecting the simultaneous presence of a bra and a ket in the expectation values. It is important to realize that the time arrow is oppositely oriented for the two members of the CTP doublet, called chronons, because they are needed to represent time reversal non-invariant interactions \cite{breakd}. It is sometime advantageous to use the parametrization $x^\pm=x\pm x^d/2$ \cite{keldysh}, $x^-$ and $x^+$ couple to the retarded and advanced Green functions, respectively, indicating the presence of the oppositely running time in this scheme. The expectation value, $\la x\ra$, can be associated with the classical coordinate and $x^d$, with $\la x^d\ra=0$, $\la x^{d2}\ra=\ord{\hbar}$ represents the quantum fluctuations \cite{effth}. 

The CTP formalism is the natural scheme to deal with the effective dynamics with non-conservative forces. Let us suppose that the observed system interacts with its environment and the closed dynamics of the full system is defined by the action $S[x,y]=S_s[x]+S_e[x,y]$, where $y$ denotes the environment coordinates. The reduced density matrix,
\be\label{effgen}
\rho(x^+_f,x^-_f)=\sum_n\la x^+_f|\otimes\la n|U(t_f,t_i)\rho_iU^\dagger(t_f,t_i)|n\ra\otimes|x^-_f\ra,
\ee
where the sum is over an environment basis, can be written as a CTP path integral,
\be\label{reddm}
\rho(x^+_f,x^-_f)=\int D[\hx]D[\hy]e^{\ih S[x^+,y^+]-\ih S[x^-,y^-]+\ih S_\epsilon[\hx]+\ih S_\epsilon[\hy]},
\ee
where the integration is over the same system trajectories as in the right hand side of eq. \eq{densm}, and the environment trajectories, $\hy(t)=(y^+(t),y^-(t))$, $t_i<t<t_f$, satisfy $y^+(t_f)=y^-(t_f)$. The convolution with the initial density matrix, shown explicitly in eq. \eq{densm}, is suppressed for the sake of the easier readability of the equations. The bare effective action, $S_{eff}[\hx]$, is introduced by writing eq. \eq{reddm} in the form
\be\label{effreddm}
\rho(x^+_f,x^-_f)=\int D[\hx]e^{\ih S_{eff}[\hx]+\ih S_\epsilon[\hx]}.
\ee
The effective action is of the form $S_{eff}[\hx]=S_s[x^+]-S_s[x^-]+S_{infl}[\hx]$, where the influence functional \cite{feynman} is given by
\be\label{qinfl}
e^{\ih S_{infl}[\hx]}=\int D[\hy]e^{\ih S_e[x^+,y^+]-\ih S_e[x^-,y^-]+\ih S_\epsilon[\hy]}.
\ee
Though we assume, for the sake of simplicity, that the initial density matrix is factorisable as the product of system and environment factors one can obviously use this scheme for entangled initial states, too. The dynamically different terms in $S_{eff}$ are better separated by using the form
\be\label{effactsep}
S_{eff}[\hx]=S_1[x^+]-S_1^*[x^-]+S_2[\hx],
\ee
where $\delta^2S_2[\hx]/\delta x^+\delta x^-\ne0$. The single trajectory action, $S_1$, describes the closed, conservative part of the effective dynamics. The chronon coupling, $S_2$, owes its existence to the presence of several non-vanishing contributions to the sum in eq. \eq{effgen}, it represents the mixed state contributions to the reduced density matrix and makes the effective system dynamics open, non-conservative. If $\Im S_1\ne0$ then the excitations, described by $S_1$, have finite life-time and the unitarity of the time evolution is preserved by the help of $S_2$. The decoherence in the coordinate diagonal representation, the suppression of the contributions with large $x^+(t)-x^-(t)$, is driven by $\Im S_2$.

\section{An effective theory}\label{effths}
A simple effective Lagrangian of a one-dimensional particle, interacting with a gas and moving under a potential $V(x)$, derived in the $\ord{x^2}$ and $\ord{\partial_t^2}$ order is
\be\label{dlagr}
L_{eff}=m\dot x^d\dot x-V\left(x+\frac{x^d}2\right)+V\left(x-\frac{x^d}2\right)-kx^d\dot x+\frac{i}2(d_0x^{d2}+d_2\dot x^{d2})
\ee
and the effective parameters depend on the microscopic details of the gas dynamics \cite{gas}. The Euler-Lagrange equation, corresponding to $x^d$,
\be\label{eeom}
m\ddot x=-V'(x)-k\dot x+i(d_0x^d-id_2\ddot x^d),
\ee
allows us to identify $k$ with the friction coefficient. The decoherence is generated by the last two terms on the right hand side of eq. \eq{dlagr}. Note that if the coordinate of a harmonic oscillator, $V(x)=m\omega_0^2x^2/2$, is considered as a Fourier component of a quantum field then the $\ord{\partial_t^0}$ part of the Lagrangian represents a complex mass, $M^2=\omega_0^2-id_0/m$.

One can easily obtain the equation of motion for the density matrix by the infinitesimal increase of the final time in eq. \eq{effreddm}, described by the master equation \cite{gas},
\be\label{master}
\partial_t\rho=\frac1{i\hbar}[H,\rho]+\left[-\frac1{2\hbar}\left(d_0+\frac{d_2k^2}{m^2}\right)x^{d2}-\frac{k}mx^d\nabla_{x^d}+\frac{id_2k}{m^2}x^d\nabla_x+\frac{d_2\hbar}{2m^2}\nabla^2_x\right]\rho,
\ee
with $H=p^2/2m+m\omega_0^2x^2/2$. Another, perhaps more illuminating form of this equation is
\be\label{wignerwv}
\dot\rho=\left[-\frac1{m}p_k\nabla_{x^d}-\frac{i}\hbar V\left(x+\frac{x^d}2\right)+\frac{i}\hbar V\left(x-\frac{x^d}2\right)-\frac{d_0}{2\hbar}x^{d2}-\frac{d_2}{2\hbar m^2}p_k^2\right]\rho,
\ee
where $p_k=p+kx^d$ with $p=-i\hbar\nabla_x$. The contribution $i\hbar\nabla_{x^d}\nabla_x/m$ on the right hand side together with the terms containing the potential $V$ reproduce the Neumann equation, the first term on the right hand side of eq. \eq{master}. The last two terms, proportional to $d_0$ and $d_2$, generate decoherence. The friction is represented by the term $kx^d$ in shifted momentum, $p_k$. These equations preserve the probability and keep the density matrix Hermitian and positive \cite{gas} for
\be\label{positin}
\nu<2\sqrt{\frac{d_0d_2}{m(m+4d_2)}}.
\ee

\section{Propagator in the Liouville space}\label{propls}
The full dynamics of a closed system is captured by the propagator \eq{trampl}. In the case of an open systems this role is taken over by the matrix element of the time evolution operator for the density matrix in the Liouville space,  $\la\hx_f|{\cal U}_t|\hx_i\ra$. To arrive at an analytical solution we restrict our attention to a harmonic oscillator where the path integral, \eq{reddm}, is easy to find,
\be\label{propdmho}
\la\hx_f|{\cal U}_t|\hx_i\ra={\cal N}_te^{\ih S_t(\hx_f,\hx_i)},
\ee
$S_t(\hx_f,\hx_i)$ denoting the CTP action, evaluated on the trajectories $\hx(t)$ which satisfy the equations of motion and the boundary conditions $\hx(t_i)=\hx_i$ and $\hx(t_f)=\hx_f$. 

The frequency spectrum is discrete for $t=t_f-t_i<\infty$ and the limit $\epsilon\to0$ can be performed already at the level of the equations of motion. Thus our system of equations to solve, the Euler-Lagrange equations of the Lagrangian \eq{dlagr}, is
\bea\label{eoms}
0&=&-\ddot x-\omega^2_0x-\nu\dot x+\frac{i}m(d_0x^d-d_2\ddot x^d),\nn
0&=&-\ddot x^d-\omega^2_0x^d+\nu\dot x^d,
\eea
where the notation $\nu=k/m$ is introduced. The solution we use to evaluate $S_t(\hx_f,\hx_i)$ is of the form
\be
\hx(t)=\sum_{\sigma,\sigma'=\pm}c_{\sigma,\sigma'}(\hx_i,\hx_f)e^{-i\omega_{\sigma,\sigma'}t},
\ee
where the coefficients $c_{\sigma,\sigma'}(\hx_i,\hx_f)$ are determined by the boundary conditions and $\omega_{\sigma,\sigma'}=\sigma\omega_\nu+i\sigma'\nu/2$ with $\omega_\nu=\sqrt{\omega_0^2-\nu^2/4}$. The frequencies $\omega_{+,-}$, $\omega_{-,-}$ belong to the spectrum of the classical, damped oscillator and the remaining exponentially diverging modes, $\omega_{+,+}$, $\omega_{-,+}$, are due to the ``wrong'' sign of the friction term in the second equation in \eq{eoms} and reflect the presence of opposite time arrows in this scheme.

The resulting action can conveniently be parametrized by the renormalized action,
\bea\label{st}
S_t(\hx_f,\hx_i)&=&\frac{M}t(x_f-x_i)(x_f^d-x_i^d)-t\frac{M\Omega^2}4(x_f+x_i)(x^d_f+x^d_i)\nn
&&-\frac{K+L}2(x_f^d+x_i^d)(x_f-x_i)-\frac{K-L}2(x_f^d-x_i^d)(x_f+x_i)\nn
&&+i\frac{D_0t}8(x^d_i+x_f^d)^2+i\frac{D_2}{2t}(x^d_f-x_i^d)^2-i\frac{D}2(x^d_i+x_f^d)(x^d_f-x_i^d)
\eea
whose coefficients can be expressed in terms of the parameters of the Lagrangian,
\be\label{runnuningparse}
M=m\omega_\nu te^{-\frac{\nu}2t}\frac{(e^{\frac\nu2t}+1)^2\cos^2\frac{\omega_\nu t}2+(e^{\frac\nu2t}-1)^2\sin^2\frac{\omega_\nu t}2}{4\sin\omega_\nu t},
\ee
\be
\Omega^2=\frac4{t^2}\frac{(e^{\frac\nu2t}-1)^2\cos^2\frac{\omega_\nu t}2+(e^{\frac\nu2t}+1)^2\sin^2\frac{\omega_\nu t}2}{(e^{\frac\nu2t}+1)^2\cos^2\frac{\omega_\nu t}2+(e^{\frac\nu2t}-1)^2\sin^2\frac{\omega_\nu t}2},
\ee
\be
K=\frac{m\nu}2,
\ee
\be
L=\frac{m\omega_\nu}{2\sin\omega_\nu t}(e^{\frac{\nu}2t}-e^{-\frac{\nu}2t}),
\ee
\bea
D_0&=&\frac{\omega_\nu}{2\omega_0^2\nu t}\frac{(e^{\frac{\nu}2t}\cos\omega_\nu t-1)^2+e^{\nu t}\sin^2\omega_\nu t}{\sin^2\omega_\nu t}\nn
&&\times[(d_0+d_2\omega_0^2)\omega_\nu(1-e^{-\nu t})+(d_0-d_2\omega_0^2)\nu e^{-\frac\nu2t}\sin\omega_\nu t],
\eea
\bea
D_2&=&\frac{\omega_\nu t}{8\omega_0^2\nu}\frac{(e^{\frac{\nu}2t}\cos\omega_\nu t+1)^2+e^{\nu t}\sin^2\omega_\nu t}{\sin^2\omega_\nu t}\nn
&&\times[(d_0+d_2\omega_0^2)\omega_\nu(1-e^{-\nu t})+(d_2\omega_0^2-d_0)\nu e^{-\frac\nu2t}\sin\omega_\nu t],
\eea
and
\be\label{runnuningparsu}
D=\frac{d_0+d_2\omega_0^2}{8\omega_0^2\nu}\frac{2\omega_\nu^2(e^{-\nu t}+e^{\nu t})+\nu^2\cos2\omega_\nu t-4\omega_0^2}{\sin^2\omega_\nu t}.
\ee
The dynamics, induced by the Lagrangian \eq{dlagr} preserves the total probability \cite{gas} and therefore
\be
{\cal N}_t=\frac1{2\pi\hbar}\left(\frac{M}t+\frac{M\Omega^2t}4+\frac{K+L}2\right).
\ee

The renormalized trajectory starts at $t=0$ at the vale of the parameters of the bare Lagrangian, keeps the decoherence parameters positive, $D,D_0,D_2>0$ and each coefficient in the renormalized action, \eq{st}, develops singularities at $t_n=n\pi/\omega_\nu$, a remnant of the focusing and anti-focusing, $x(t_n)=(-1)^nx_i$, of the undamped classical harmonic oscillator  \cite{rgqm}.

\section{Gaussian Wave packet}\label{waveps}
We use now the Liouville space propagator to find the time evolution of a Gaussian pure wave packet,
\be
\psi_i(x)=\sqrt{\frac{2\sqrt{\pi\hbar}}\kappa}\int\frac{dq}{2\pi}e^{iqx-\frac\hbar{2\kappa^2}q^2}.
\ee

\subsection{Density matrix}
The convolution of the propagator in the Liouville space \eq{propdmho} with the initial density matrix is a Gaussian integral yielding
\be\label{gdensm}
\rho(x,x^d)=\frac1{\sqrt{2\pi}\sigma_x}\exp\left[-\frac{x^2}{2\sigma_x^2}-\frac{x^{d2}}{2\sigma_{xd}^2}+i\frac{xx^d}{\ell_{xd}^2}\right].
\ee
The decoherence may show up in mixed states only. The Gaussian density matrices can be written in the form
\be
\rho(x^+,x^-)=\frac1{\sqrt{2\pi}\sigma_x}e^{-\hf(\frac1{4\sigma^2_x}+\frac1{\sigma^2_{xd}}-\frac{i}{\ell^2_{xd}})x^{+2}-\hf(\frac1{4\sigma^2_x}+\frac1{\sigma^2_{xd}}+\frac{i}{\ell^2_{xd}})x^{-2}+(\frac1{\sigma^2_{xd}}-\frac1{4\sigma^2_x})x^+x^-},
\ee
showing that the distance of the purity
\be\label{purity}
\gamma=\Tr\rho^2=\frac{\sigma_{xd}}{2\sigma_x},
\ee
from $1$ characterizes the mixing and is a measure of the decoherence. 

It is important to know the density matrix in momentum space, 
\be
\rho(p,p^d)=\int dxdx^d\rho(x,x^d)e^{-\ih(x+\frac{x^d}2)(p+\frac{p^d}2)+\ih(x-\frac{x^d}2)(p-\frac{p^d}2)},
\ee
which assumes the form
\be\label{gdensmp}
\rho(p,p^d)=\frac{\sqrt{2\pi}\hbar}{\sigma_p}\exp\left[-\frac{p^2}{2\sigma_p^2}-\frac{p^{d2}}{2\sigma^2_{pd}}+i\frac{pp^d}{\pi_{pd}^2}\right],
\ee
with
\bea
\sigma^2_p&=&\hbar^2\left(\frac{\sigma_x^2}{\ell_{xd}^4}+\frac1{\sigma_{xd}^2}\right),\nn
\sigma^2_{pd}&=&\hbar^2\left(\frac{\sigma_{xd}^2}{\ell_{xd}^4}+\frac1{\sigma_x^2}\right),\nn
\pi^2_{pd}&=&-\hbar^2\frac{1+\frac{\ell_{xd}^4}{\sigma_x^2\sigma_{xd}^2}}{\ell_{xd}^2}.
\eea
Note that the basis (gauge) transformation $\psi(x)\to e^{i\frac\lambda2x^2}\psi(x)$ acts on the density matrix \eq{gdensm} as $1/\ell_{xd}^2\to1/\ell_{xd}^2-\lambda$. Therefore the imaginary part of $\rho(x,x^d)$ and $\rho(p,p^d)$ can be canceled by $\lambda=1/\ell_{xd}^2$ and the density matrix, obtained in such a manner, belongs to a representation where the momentum operator is $p=-i\hbar\nabla_x+\hbar\lambda x$.

A Gaussian wave packet has three important parameters, namely the position and the width of the probability distribution of the coordinate, $x$, and the width of the quantum fluctuations, $x^d$. The position of the peak is governed by the classical equation of motion, reflects classical physics only and will be ignored in what follows. The widths, the coefficients of the $\ord{x^2}$ and $\ord{p^2}$ terms in the exponent represent genuine quantum effects, such as the spread of the wave packet and the decoherence. 

The parameters of the density matrix are found by using the parametrization \eq{st} in the propagator \eq{propdmho},
\be\label{paramx}
\sigma^2_x=\hbar\frac{N_x}{D_x},~~~
\sigma^2_{xd}=\hbar\frac{N_x}{D_d},~~~
\ell^2_{xd}=\hbar\frac{N_x}{D_{xd}},
\ee
where
\bea\label{runnpar}
N_x&=&4Mt(L-K)(\Omega^2t^2-4)+M^2(\Omega^2t^2-4)^2+4t^2(K-L)^2\nn
&&+8t\kappa^2[4D_2+t(4D+D_0t+2\hbar\kappa^2)],\nn
D_x&=&2\kappa^2[2(K+L)t+M(\Omega^2t^2+4)]^2,\nn
D_d&=&4D_0t(2M+Lt)^2+\kappa^2\{2[2M-(K+L)t]^2,\nn
&&+4M^2\Omega^2t^2+\hf M^2\Omega^4t^4-2(K+L)M\Omega^2t^3-32D^2t^2+4D_0\kappa^2t^3\}\nn
&&+4D_2t[K^2-4KM\Omega^2t+M^2\Omega^4t^2+4\kappa^4(2D_0t+\kappa^2)]\nn
&&-8Dt(4MK-2LtK+LM\Omega^2t^2-2M^2\Omega^2t+2\kappa^4t),\nn
D_{xd}&=&4t[2K^2Lt+2LM^2\Omega^2t-M^3\Omega^2(4-\Omega^2t^2)+2Lt\kappa^2(2D+D_0t+\kappa^2)\nn
&&+M\kappa^2(8D+4D_0t-8D_2\Omega^2t-2D\Omega^2t^2+4\kappa^2-\Omega^2t^2\kappa^2)-2KL^2t\nn
&&-KLM(\Omega^2t^2-4)-2M^2K\Omega^2t-2K\kappa^2(4D_2+2Dt+t\kappa^2)].
\eea

The particle polarizes the gas and the polarization cloud follows the motion of the particle, representing an interaction induced spread of the wave packet, in addition to the usual spread, owing to the dephasing of the non-interacting particle state. The irreversibility of the effective dynamics arises from this spread. It is important to bear in mind that the irreversibility appears differently on the level of first and the second moments of the coordinate and the momentum. While the first moment displays an $\ord{\hbar^0}$ asymmetric spread, corresponding to a friction force, the symmetric part of the spread induces $\ord\hbar$ second moments.

The last step of the calculation is the insertion of the expressions  \eq{runnuningparse}-\eq{runnuningparsu} into these equations. In this process one generates a sum of ratios where the numerators and the denominators are sums of terms, each being the product of the parameters of the Lagrangian, the time and an exponential function of the form $\exp(j\nu+ik\omega_n)t/2$ with $j$ and $k$ integers. The final expressions are far too long to be reproduced here. This structure is dominated by different terms in the limit $t\to\infty$ and makes the asymptotic density matrix, reached after some transient oscillation, discontinuous when one of the parameters of the Lagrangian is sent to zero.

There are several time scales in the dynamics of the damped harmonic oscillator. The Liouville space propagator contains two time scales, $\tau_{fr}=\nu^{-1}$ corresponds to a free particle under the influence of the friction force, and $\tau_{osc}=1/\sqrt{|\omega_0^2-\nu^2/4|}$ is the characteristic time of the damped oscillator. Both time scales are classical, free of decoherence effects. The decoherence arising at a given chronon separation, $x^d$, involves another, $x^d$-dependent time scale, 
\be\label{dectime}
\tau_d(x^d)=\frac\hbar{d_0x^{d2}}
\ee
and appears extremely short at macroscopic chronon separations \cite{zurek,joos}. Note that this scaling law holds for harmonic systems only.

\subsection{Free particle}
The best is to start the exploration of the time dependence with the translation invariant case, $\omega_0=0$, the motion of a decohering particle under the influence of a friction force. The width of the state in the coordinate and the momentum space turns out to be
\bea
\sigma_x^2&=&\frac\hbar{2\kappa^2}\left[1+\frac{\kappa^2}{m^2\nu^3}[\kappa^2\nu(1-e^{-t\nu})^2+d_2\nu^2(1-e^{-2t\nu})+d_0(4e^{-t\nu}-e^{-t\nu}+2t\nu-3)]\right],\nn
\sigma_p^2&=&\frac\hbar{2\nu}\{d_0(1-e^{-2t\nu})+\nu[\kappa^2e^{-2t\nu}+d_2\nu(1-e^{-2t\nu})]\},
\eea
with the long time asymptotic forms
\be\label{sigmaxfree}
\sigma_x^2\approx\frac{d_0\hbar t}{k^2}=\sigma_0^2\frac{2d_0}{\kappa^2\nu^2t},
\ee
and
\be
\sigma_p^2\approx\frac\hbar{2\nu}(d_0+d_2\nu^2),
\ee
where $\sigma^2_0=\hbar\kappa^2t^2/2m^2$ stands for the width of the free particle state. The lesson of this result is the following: (i) The spread of the state is slowed down in the coordinate space compared to the free particle free of decoherence. This is expected on the ground that the friction slows down the motion. (ii) The widths are  independent of the initial state, suggesting the existence of an attractive IR fixed point in the time evolution, a relaxed state. (iii) Both widths are reduced by the friction.  This seems rather natural, as well, since the friction reduces the mobility and suppresses the higher momentum components of the state. However, it raises the possibility of a conflict with the uncertainty principle, a problem to be addressed below. (iv) The dephasing, the dynamical origin of the spread of the wave packet, is enhanced by the decoherence. (v) In the effective theory of a test particle, moving in a gas, the perturbative derivation of the parameters of the effective Lagrangian \eq{dlagr} yields $\ord{g^2}$ contributions to the parameters $\nu$, $d_0$ and $d_2$ where $g$ denotes the coupling constant of the test particle to its environment \cite{gas}. Therefore $\sigma_p^2=\ord{g^0}$ is a nontrivial width of the asymptotic state in the momentum space, formed even by an infinitesimal interaction. Such a persistent effect of weak interactions accumulates due to the conservation of the momentum.

The decoherence strengths, $\sigma_{xd}^2=\hbar\nu N_{xd}/D_d$ and $\sigma_{pd}^2=\hbar N_{pd}/D_d$ are determined by
\bea
N_{xd}&=&\nu[\kappa^4(e^{t\nu}-1)^2+d_2\kappa^2\nu(e^{2t\nu}-1)+k^2e^{2t\nu}]+d_0\kappa^2[4e^{t\nu}-1+e^{2t\nu}(2t\nu-3)]\nn
N_{pd}&=&2\kappa^2k^2\nu[d_0(e^{2t\nu}-1)+\nu[\kappa^2+d_2\nu(e^{2t\nu}-1)]\nn
D_d&=&d_0^2(e^{t\nu}-1)\kappa^2[t\nu+2+e^{t\nu}(t\nu-2)]+\frac{\nu^3}2[mk\kappa^2+d_2(\kappa^4+k^2)(e^{2t\nu}-1)]\nn
&&+\frac{d_0\nu}2[(e^{2t\nu}-1)k^2+2d_2\kappa^2\nu^2t(e^{2t\nu}-1)+\kappa^4(3-4e^{t\nu}+e^{2t\nu}+2t\nu)],
\eea
and their asymptotic forms are 
\be
\sigma_{xd}^2\approx\hbar\frac{2\nu}{d_0+d_2\nu^2},
\ee
and 
\be
\sigma_{pd}^2\approx\hbar\frac{\nu^2m^2}{d_0t}.
\ee
One learns here: (i) The decoherence parameters of the effective Lagrangian, $d_0$ and $d_2$, indeed generate decoherence. (ii) The friction tends to recohere the particle. This is rather surprising, one would have expected the strengthening, rather than the weakening of the decoherence with the friction constant, this latter being a coupling constant to the  environment. Though the origin of the effective parameters $\nu$ and $d_j$ is common, the leakage of the particle state to the environment, their role in forming the decoherence in the coordinate and the momentum space is the opposite. (iii) Though the decoherence becomes complete in the the momentum space owing to the momentum conservation, there is some finite coherence left in the the coordinate space. (iv) $\sigma_{xd}^2=\ord{g^0}$, a partial decoherence in coordinate space is achieved even for infinitesimal interaction strength.

Together with the remaining parameters,
\bea
\ell_{xd}^2&=&-\hbar\frac{\nu[\kappa^4(e^{t\nu}-1)^2+d_2\kappa^2\nu(e^{2t\nu}-1)+e^{2t\nu}k^2]+d_0\kappa^2[4e^{t\nu}-1+e^{2t\nu}(2t\nu-3)]}{k\kappa^2(e^{t\nu}-1)\{d_0(e^{t\nu}-1)+\nu[d_2\nu(e^{t\nu}+1)-\kappa^2]\}}\nn
&\approx&-\hbar\frac{2d_0t}{m(d_0+d_2\nu^2)}\nn
\pi_{pd}^2&=&\hbar\nu m\frac{d_0(e^{2t\nu}-1)+\nu[\kappa^2+d_2\nu(e^{2t\nu}-1)]}{(e^{t\nu}-1)\{\nu[d_2\nu(e^{t\nu}+1)-\kappa^2]-d_0(e^{t\nu}-1)\}}\nn
&\approx&\hbar\nu m\frac{d_2\nu^2+d_0}{d_2\nu^2-d_0}
\eea
one finds the density matrix,
\be\label{asdensx}
\rho(x,x^d)\approx\frac{m\nu}{\sqrt{2\pi d_0\hbar t}}e^{-\frac{m^2\nu^2}{2d_0\hbar t}x^2-\frac{d_0+d_2\nu^2}{4\nu\hbar}x^{d2}-i\frac{m(d_0+d_2\nu^2)}{2d_0\hbar t}x^dx},
\ee
in the coordinate space purity, describing a state which spreads into a translation invariant shape with vanishing purity. The density matrix relaxes to a completely decohered Gibbs operator with temperature $k_BT=\hbar(d_0+d_2\nu^2)/2m\nu$ for the relaxed state,
\be\label{freeasdp}
\rho(p,p^d)\approx\sqrt{\frac\nu{\pi\hbar(d_0+d_2\nu^2)}}e^{-\frac\nu{\hbar(d_0+d_2\nu^2)}p^2-\frac{d_0t}{2\hbar m^2\nu^2}p^{d2}+i\frac{d_2\nu^2-d_0}{\hbar\nu m(d_2\nu^2+d_0)}p^dp},
\ee
in momentum space. Both density matrices assume a nontrivial form even for infinitesimally weak system-environment interactions. The time dependence of a Gaussian density matrix of the free particle has already been explored by the help of a master equation, containing the $\ord{d_0}$ term only in the square bracket of the right hand side of eq. \eq{master} \cite{joos,morikawa} and $\sigma_x^2=\ord{t^3}$ and $\sigma_{xd}^2=\ord{t^{-1/2}}$ was found as $t\to\infty$. It was mentioned after eq. \eq{runnpar} that the renormalized parameters, shown in that equations, can be written as the sum of different ratios. The asymptotic, long time limit of these ratios displays different functional forms of the parameters of the effective Lagrangian and changes discontinuously when some of the parameters are canceled. Hence the comparison of the solution of the dynamics, based on different truncation of the master equation \eq{master} is not trivial. Nevertheless it is reasonable to hold the friction force responsible for both the slower increase of the width with time, $\sigma_x^2=\ord{t}$, c.f. \eq{sigmaxfree} and the finite decoherence of the asymptotic density matrix \eq{asdensx}.

\begin{figure}
\includegraphics[scale=.8]{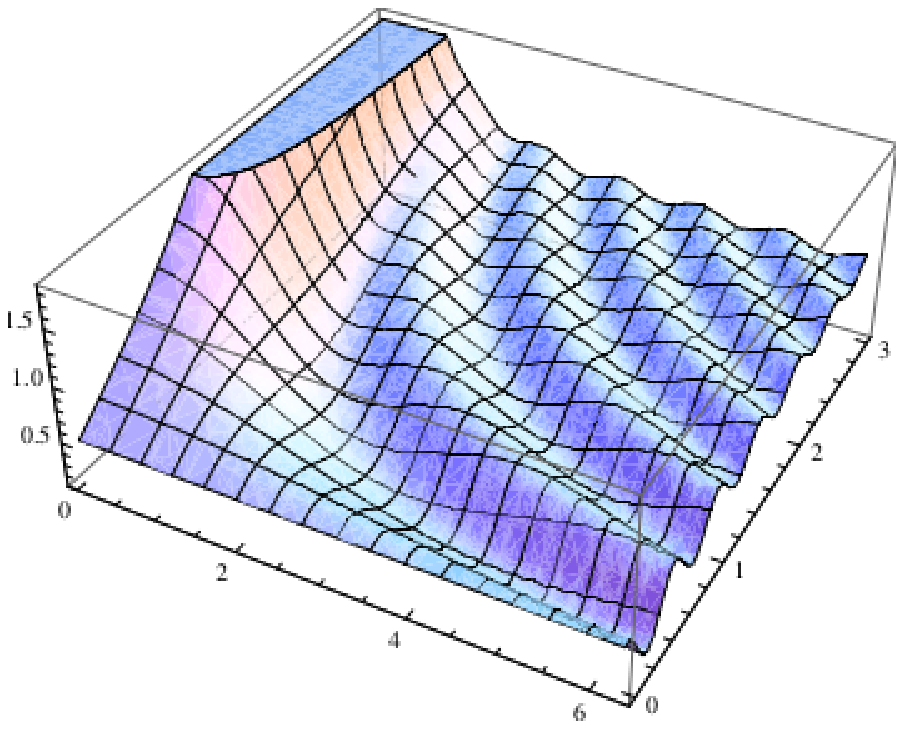}\hskip1cm\includegraphics[scale=.8]{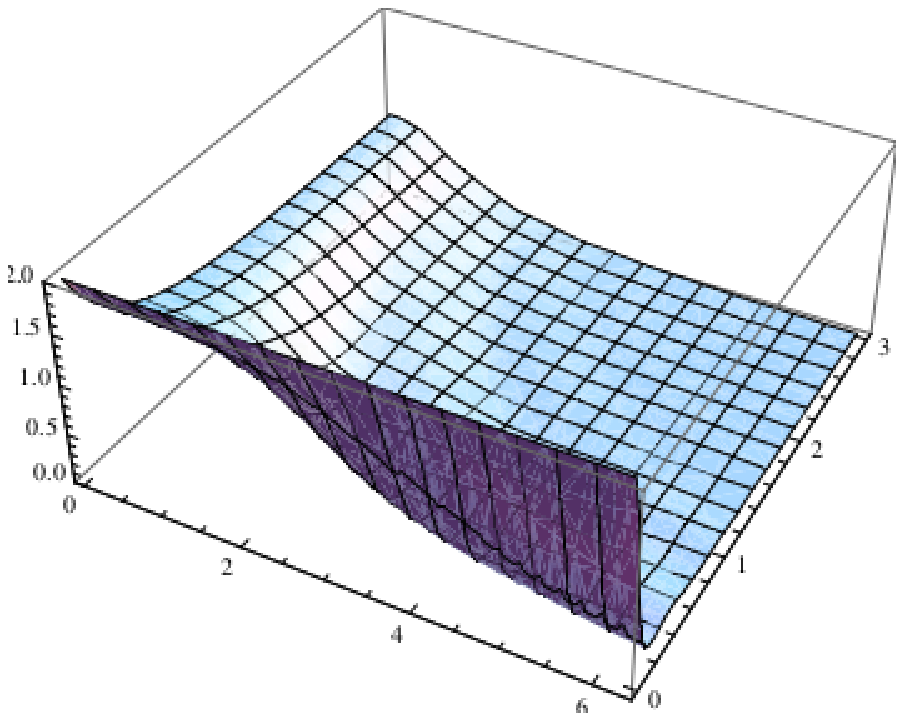}	
\centerline{(a)\hskip8.1cm(b)}

\includegraphics[scale=.8]{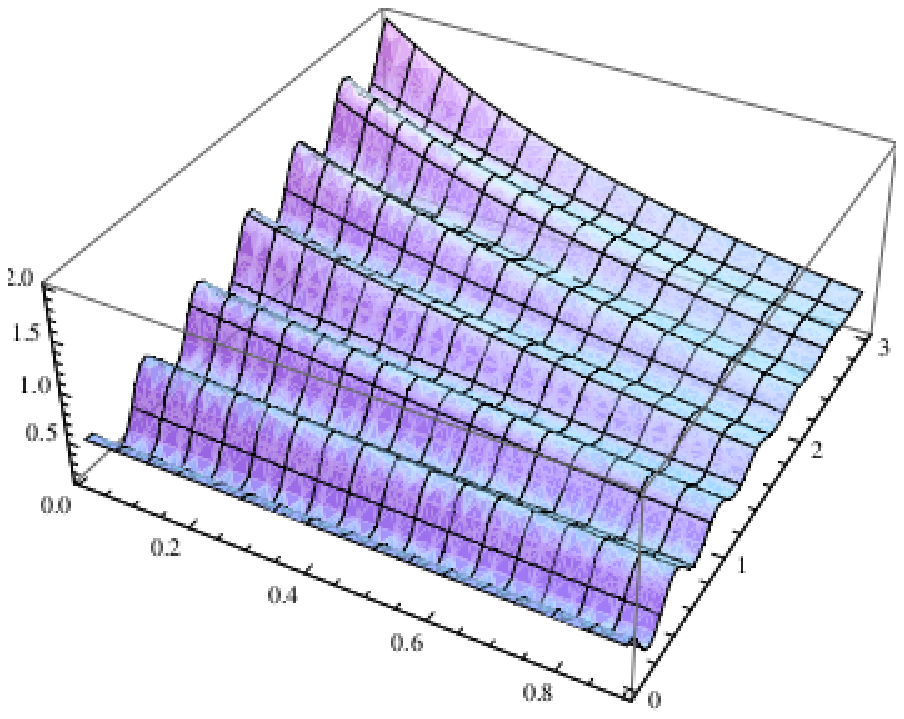}\hskip1cm\includegraphics[scale=.8]{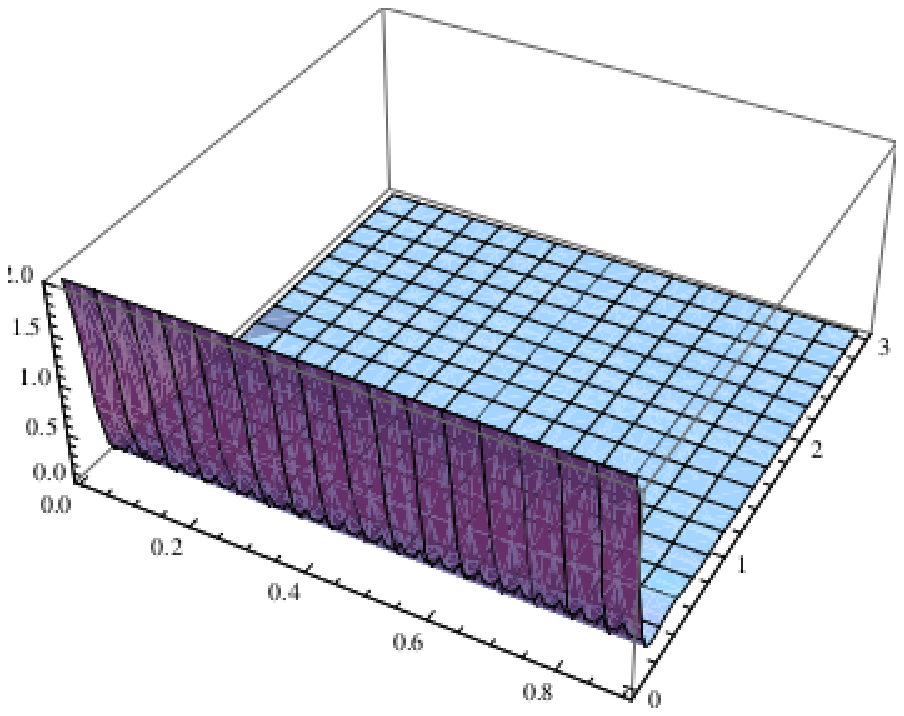}
\centerline{(c)\hskip8.1cm(d)}
\caption{$\sigma^2_x/\ell^2$ ((a), (c)) and $\sigma^2_d/\ell^2$ ((b), (d)), plotted against the $(\omega_0\tau,t/\tau)$ plane for $\nu\tau=0.5$ ((a), (b)), on the $(\nu\tau,t/\tau)$ plane for $\omega_0\tau=2\pi$ ((c), (d)). The covered $\nu$-interval is the one, allowed according to the inequality \eq{positin}.}\label{sigmax}
\end{figure}

\subsection{Time dependence for a harmonic oscillator}
The time dependence of the state of a harmonic oscillator is far more complex and we present here a few typical numerical results only, expressed in the natural mass, length and time units $\mu=m$, $\ell=\sqrt\hbar/\kappa$ and $\tau=m/\kappa^2$. The values $d_0=\mu/\tau^2$, $d_2=\mu$ were used in the numerical results, to be described below. The time dependence of the width in the coordinate space, $\sigma_x$, depicted in Fig. \ref{sigmax} (a), is oscillatory if $\nu<2\omega_0$ and becomes a monotonous function of the time for an over damped oscillator, $\nu>2\omega_0$. Furthermore it diverges as $\omega_0\to0$ and $t\to\infty$, as expected for the spread of the wave packet of a free particle. Fig. \ref{sigmax} (c) displays the width as the function of the time and the friction frequency, $\nu=k/m$, for an under damped oscillator and shows the slowing down of the spread by the friction force. One finds  qualitatively similar results for the over damped oscillator except that the time dependence is monotonic, without oscillation. The coherence of the initial state is rapidly lost in the coordinate space according to Fig. \ref{sigmax} (b).  Fig. \ref{sigmax} (d) shows that the relaxation of the decoherence to its asymptotic form is approximately independent of the friction force. 

\begin{figure}
\includegraphics[scale=.8]{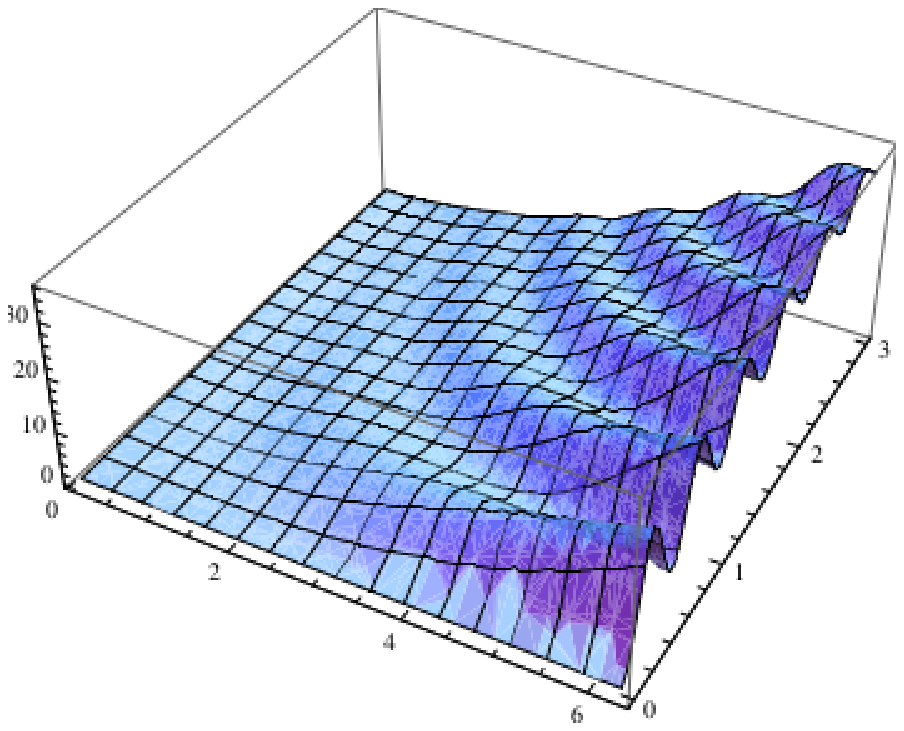}\hskip1cm\includegraphics[scale=.8]{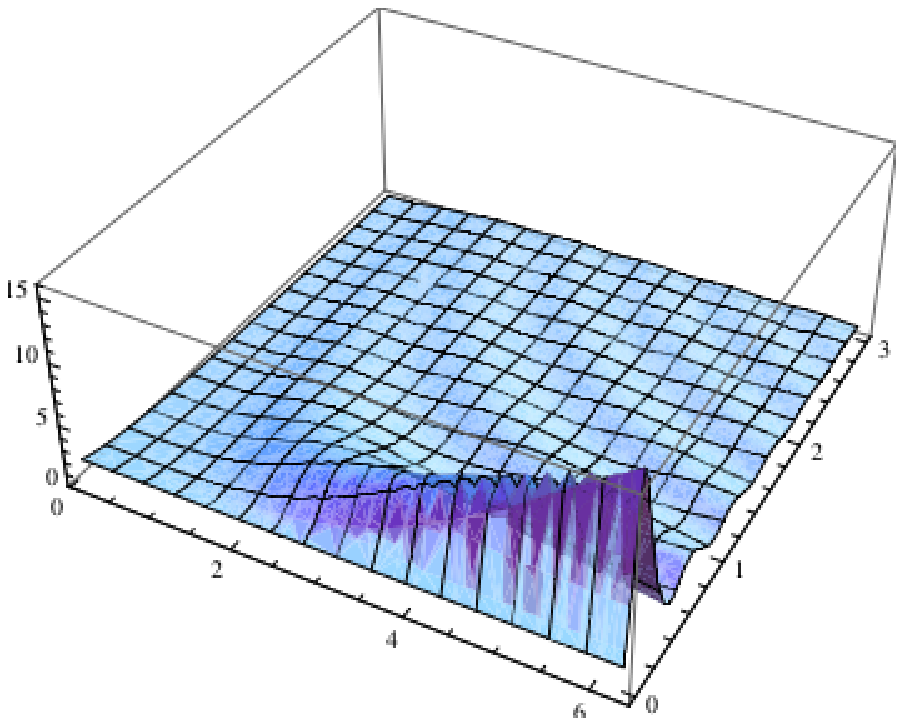}
\centerline{(a)\hskip8.1cm(b)}

\includegraphics[scale=.8]{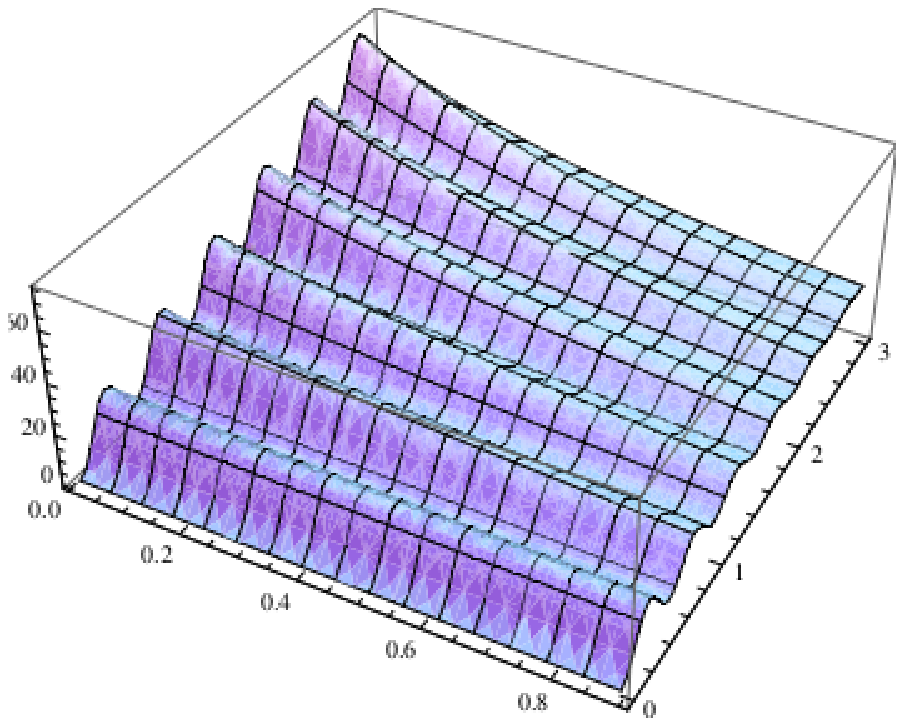}\hskip1cm\includegraphics[scale=.8]{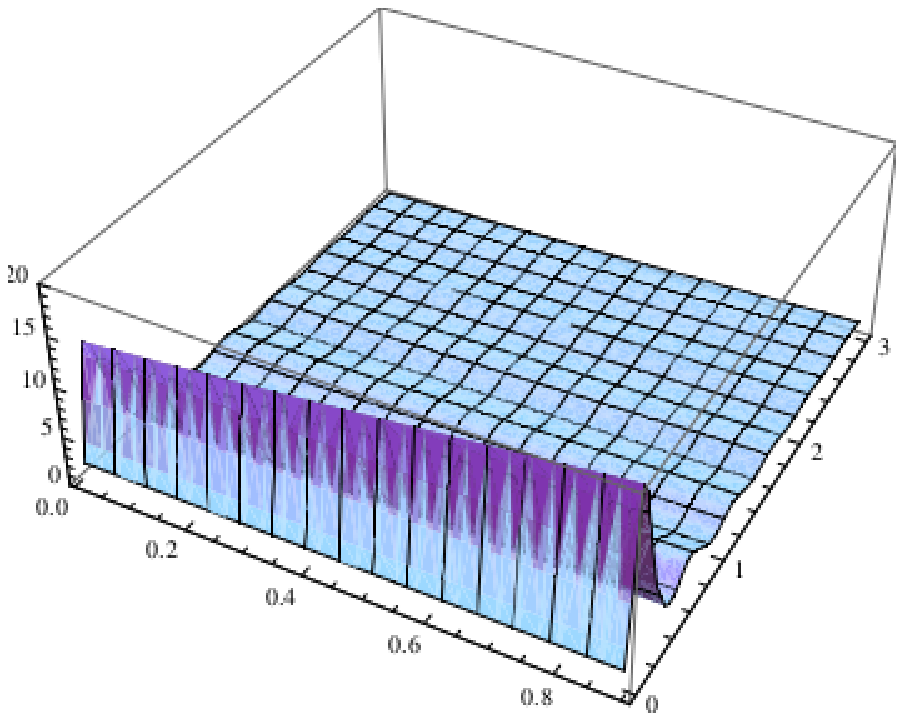}
\centerline{(c)\hskip8.1cm(d)}
\caption{The same as Fig. \ref{sigmap}, now for $\sigma^2_p/\hbar\kappa^2$ and $\sigma^2_{pd}/\hbar\kappa^2$.}\label{sigmap}
\end{figure}

The numerical results about the time evolution in the momentum space, presented in Figs. \ref{sigmap}, show the characteristic oscillations in $\sigma_p$ and $\sigma_{pd}$ for the under damped oscillator and the monotonic time dependence for the over damped case. The increasing localization in coordinate space with $\omega_0$, seen in Fig. \ref{sigmax} (a), is reflected here in the increase of $\sigma_p$, displayed in Fig. \ref{sigmap} (a). The friction also strengthens the localization simultaneously in the coordinate and the momentum space for $\omega_0\ne0$. The under damped oscillator is subject of a remarkable rapid, friction induced, transient recoherence, shown in Figs. \ref{sigmap} (b) and (d). This peak in time is absent for the over damped oscillator (not presented here) where the decoherence increases monotonically with time.

If $\ord{x}$ terms are introduced in the exponent of the initial density matrix \eq{gdensm} then $\la\v{x}\ra$ and $\la\v{p}\ra$ become non-vanishing. Since they follow the classical equations of motion, they approach zero as $t\to\infty$, canceling the $\ord{x}$ and $\ord{p}$ terms in the exponent of the asymptotic density matrix.

\subsection{Strongly and weakly localized states}
It is remarkable that the singularities of the running parameters \eq{runnuningparse}-\eq{runnuningparsu}, occurring at $t_n=n\pi/\omega_\nu$, are smeared by the interference due to the finite width of the initial state. It will be shown that the singularities return if the initial state is completely localized or delocalized. 

We start with the case of maximally localize initial state. The parameters $\sigma_{xd}^2$ and $\ell^2_{xd}$ are not sensitive to the spread of the wave packet and their value, corresponding to a point-like initial state, can be found by considering the density matrix
\be\label{naivelocl}
\rho(x,x^d)={\cal N}_t{e^{\ih S_t(\hx,\hx_i)}}_{|\hx_i=0},
\ee
where $S_t(\hx,\hx_i)$ is given by \eq{st} which is of the form of \eq{gdensm} with
$\sigma_x^2=\infty$, 
\be\label{locstatep}
\sigma^2_{xd}=\frac{8\hbar\omega_0^2\nu\sin^2\Omega t}{(d_0+d_2\omega_0^2)[4\omega_0^2(1-e^{-\nu t})+\nu^2(e^{-\nu t}-\cos2\Omega t)]-(d_0-d_2\omega_0^2)2\nu\Omega\sin2\Omega t},
\ee
and
\be\label{locstatepl}
\ell^2_{xd}=\frac\hbar{m(\Omega\cot\Omega t-\frac\nu2)}.
\ee
The width of the state, $\sigma_x^2$, which is infinite for $t>0$ in eq. \eq{naivelocl}, can be found by extracting the leading contribution to $\sigma_x^2$ in the limit $\kappa\to\infty$, using the first equation in eqs. \eq{paramx},
\be\label{asw}
\sigma_x^2=\frac{\hbar\kappa^2}{4m^2}e^{-t\nu}\frac{\sin^2\Omega t}{\Omega^2}.
\ee

\begin{figure}
\includegraphics[scale=0.6]{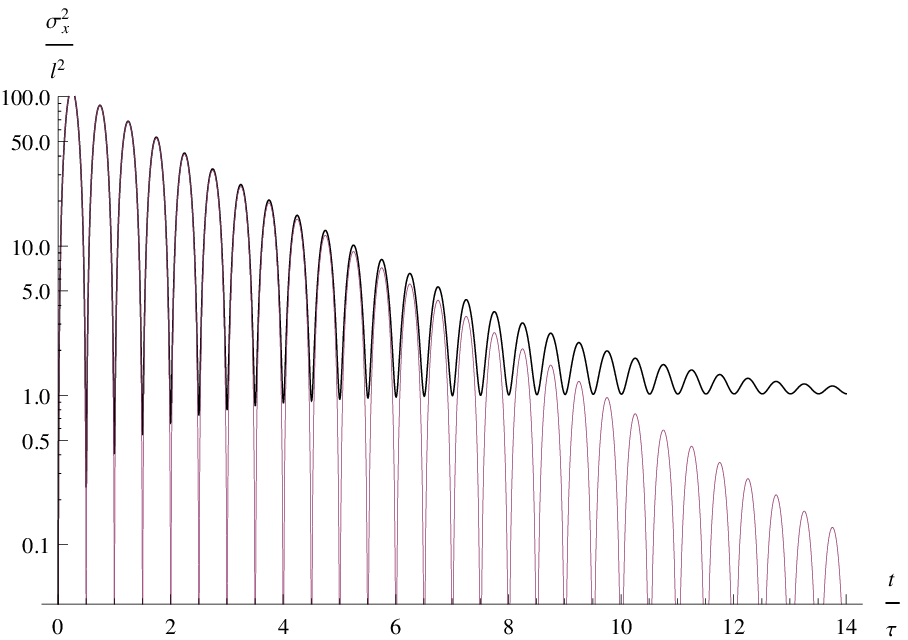}\hskip1cm\includegraphics[scale=0.6]{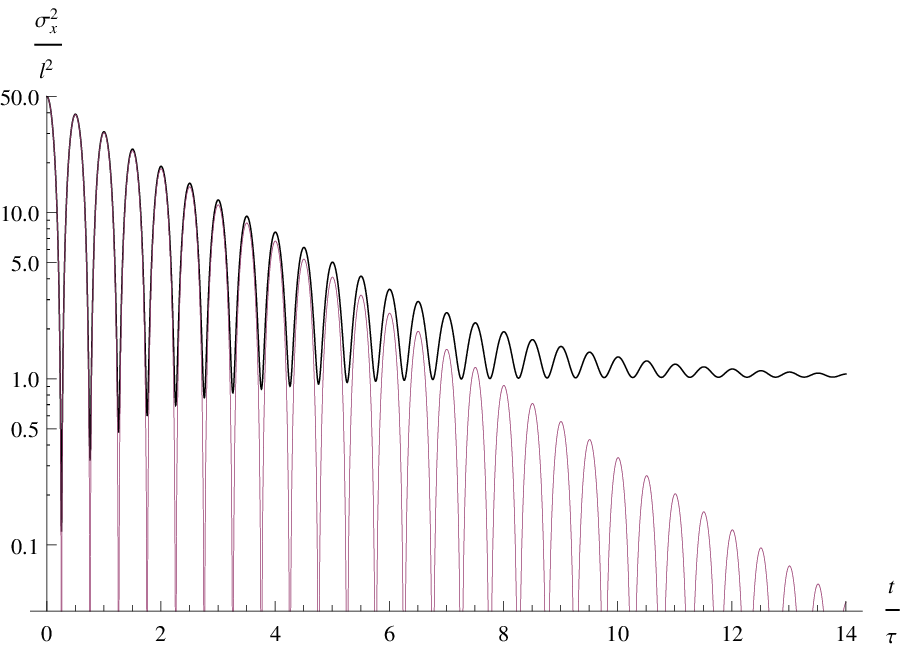}
\centerline{(a)\hskip6cm(b)}

\includegraphics[scale=0.6]{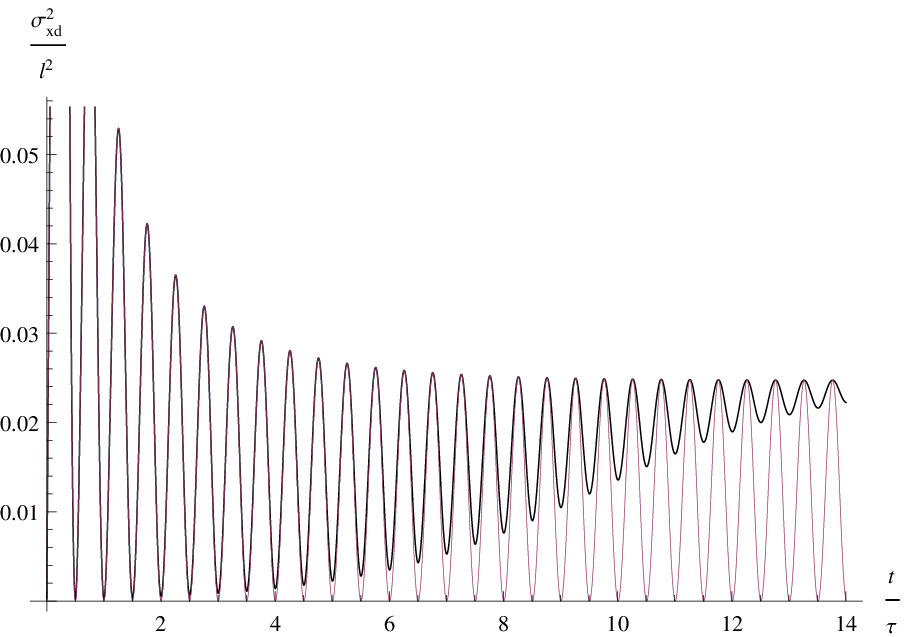}\hskip1cm\includegraphics[scale=0.6]{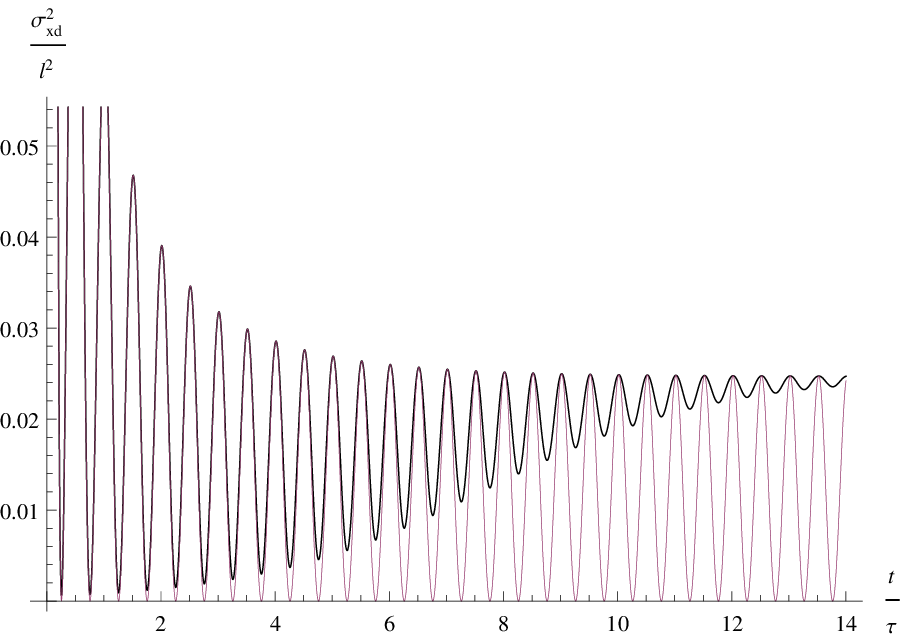}
\centerline{(c)\hskip6cm(d)}

\includegraphics[scale=0.6]{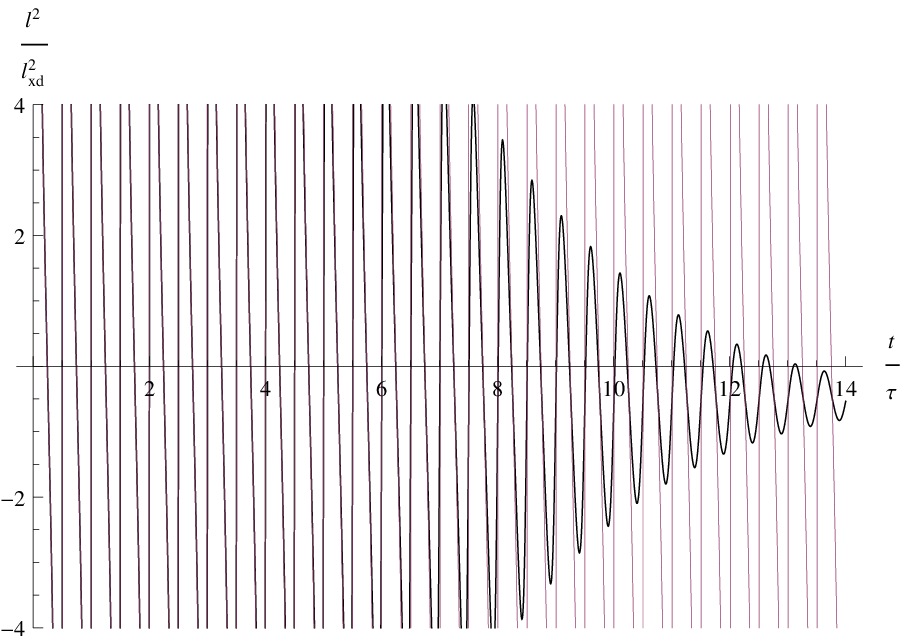}\hskip1cm\includegraphics[scale=0.6]{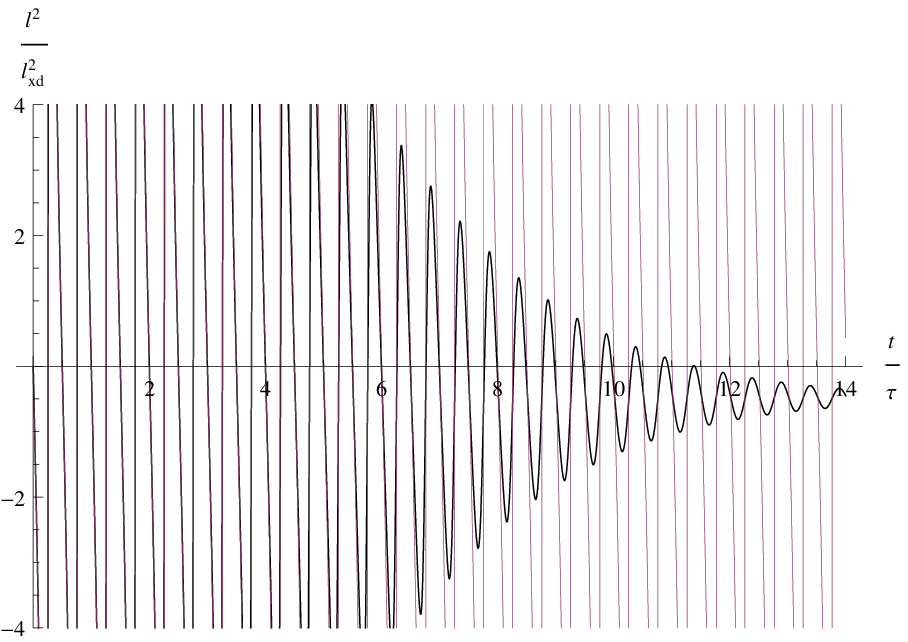}
\centerline{(e)\hskip6cm(f)}

\caption{The length parameters $\sigma^2_x/\ell^2$ (a)-(b), $\sigma^2_{xd}/\ell^2$ (c)-(d) and $\ell^2/\ell^2_{xd}$ (e)-(f) for strongly and weakly localized initial states as functions of $t/\tau$ with $\omega_0\tau=2\pi$ and $\nu\tau=0.5$. The thick lines correspond to the width parameter $\kappa\ell/\sqrt\hbar=100$ in (a), (c) and (e) and $\kappa\ell/\sqrt\hbar=0.1$ in (b), (d) and (f). The thin lines follow the asymptotic expressions.}\label{locf}
\end{figure}

The time dependence of $\sigma_x^2$, calculated for large $\kappa$ follows the asymptotic expression \eq{asw}, for a while, however, it approaches a non-vanishing limit as $t\to\infty$, as opposed to $\sigma_x^2\to0$ for $\kappa=\infty$, c.f. Fig. \ref{locf} (a). Thought the two curves of Fig. \ref{locf} (a) remain close to each other for longer time as $\kappa$ increases, nevertheless the limits $\kappa\to\infty$ and $t\to\infty$ are non-commuting. As mentioned after eq. \eq{runnpar}, $1/\sigma_x^2$ is the sum of ratios where the limit $\kappa\to\infty$, carried out at finite $t$, suppresses the terms in the numerators and the denumerators, leaving behind a friction-driven shrinking of the width. As the time passes the terms, suppressed by $1/\kappa$ become more important due to their slower decay in time and this approximation is violated. The terms, suppressed in the limit $k\to\infty$, arise from the decoherence because the time evolution with $d_0=d_2=0$ can be reproduced numerically by the asymptotic expression, \eq{asw} for arbitrarily large times. 

The other length scales of a strongly localized state, $\sigma_{xd}^2$ and $\ell_{xd}^2$, behave in a similar manner, namely they follow the oscillating asymptotic expressions, \eq{locstatep}-\eq{locstatepl}, for some time and deviate as $t\to\infty$, cf. Figs. \ref{locf} (c) and (d). The decoherence parameter, $\sigma_{xd}^2$, starts at $t=0$ with the small value of the pure initial state which is reproduced at $t_n$. However, this minimum is increasing as the peak in the parameter $1/\sigma_{xd}^2$ of the density matrix is gradually eroding with time owing to the interference within the small but finite extent initial state. This is the mechanism which drives $\sigma_{xd}^2$ to its asymptotic value as $t\to\infty$. The time dependence of the imaginary part of $\ln\rho$, parametrized by $1/\ell_{xd}^2$, is driven mainly by the kinetic energy at $t\approx t_n$. Thus one finds divergences,  $1/\ell_{xd}^2=\infty$, the remnant of the strong initial localization at $t\approx t_n$. The contributions to the momentum expectation value come from $\Im\ln\rho$, therefore $1/\ell_{xd}^2$ crosses zero around $t_{n+1/2}=(t_n+t_{n+1})/2$, at the classical turning points.

\begin{figure}
\includegraphics[scale=0.6]{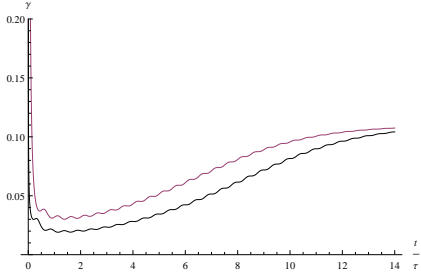}
\caption{The purity $\gamma$ as a function of $t/\tau$ for  $\kappa\ell/\sqrt\hbar=100$ (thick line) and $\kappa\ell/\sqrt\hbar=0.1$ (thin line).}\label{declocf}
\end{figure}

The purity \eq{purity}, reproduced in Fig. \ref{declocf}, shows that the strong localization, recurring at times $t_n$, is accompanied by decoherence for $n\ge1$. Furthermore, we find some recoherence after a strong transient decohering phase of the time evolution. 

The results mentioned above remain valid if the initial pure state is localized at an arbitrary space point, different form the minimum of the harmonic potential, since the quadratic part of $S_t(\hx,\hx_i)$ in $\hx$ is independent of $\hx_i$. 

It is interesting that the time dependence of a strongly delocalized state shows qualitatively similar features. In fact, the state described by an almost constant wave function in the coordinate space remains classical for a while because the quantum fluctuations, induced in the Schr\"odinger equation by the $\ord{\hbar^2}$ kinetic energy, are weak. Thus the self focusing of the classical equation of motion is expected to be recovered for a while in the time evolution. The main difference from the strongly localized case is that the peaks occur in the length scales rather than the minimas at times $t_n$, cf. Figs. \ref{locf}-\ref{declocf}.

\section{Gaussian Asymptotic state}\label{asymptss}
Though the time dependence of the density matrix can be obtained analytically the length of the expressions makes the use of the exact result rather difficult. Nevertheless we can gain some analytical insight into the impact of the friction and the decoherence upon the dynamics by looking into the asymptotic state, reached in the limit $t\to\infty$. The condition $\dot\rho=0$ is easy to fulfill for a Gaussian density matrix, yielding
\bea\label{sigmas}
\sigma^2_x&=&\frac\hbar{2m^2\nu}\frac{d_0+d_2\omega_0^2}{\omega_0^2},\nn
\sigma^2_{xd}&=&2\hbar\nu\frac{d_0+d_2\omega_0^2}{(d_0+d_2\omega_0^2)^2+d_0d_2\nu^2},\nn
\ell^2_{xd}&=&-\frac\hbar{m\nu}\frac{d_0+d_2\omega_0^2}{d_2\omega_0^2},
\eea
and
\bea\label{sigmasp}
\sigma_p^2&=&\frac\hbar{2\nu}[d_0+d_2(\nu^2+\omega_0^2)],\nn
\sigma^2_{pd}&=&2m^2\hbar\nu\omega_0^2\frac{d_0+d_2(\nu^2+\omega_0^2)}{(d_0+d_2\omega_0^2)^2+d_0d_2\nu^2},\nn
\pi_{pd}^2&=&\frac{\hbar m}\nu\frac{d_0+d_2(\nu^2+\omega_0^2)}{d_2}.
\eea

\begin{figure}
\includegraphics[scale=.8]{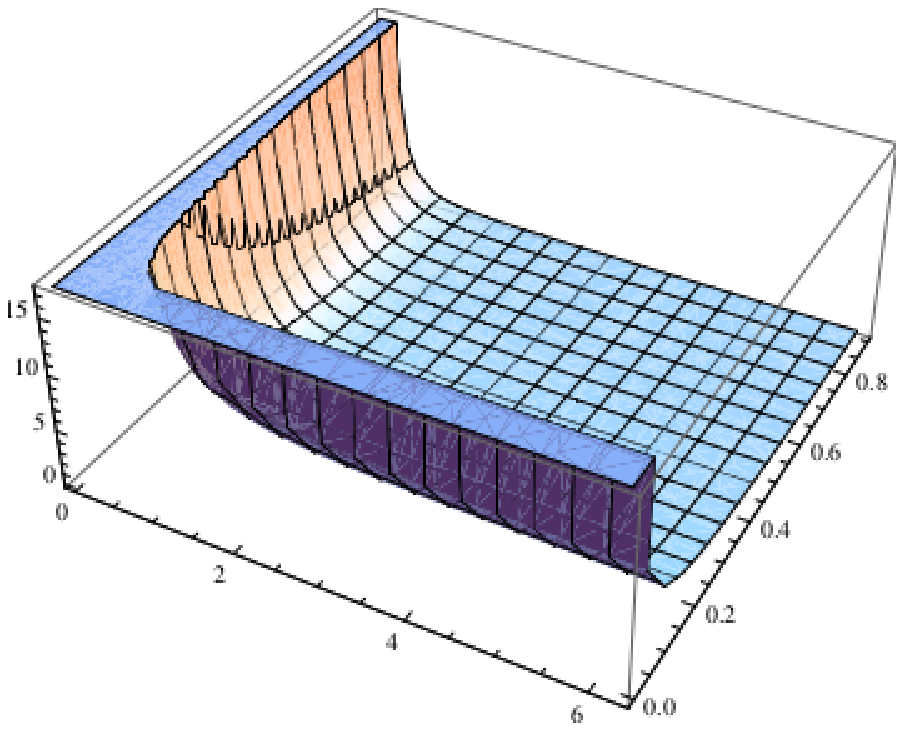}\hskip1cm\includegraphics[scale=.8]{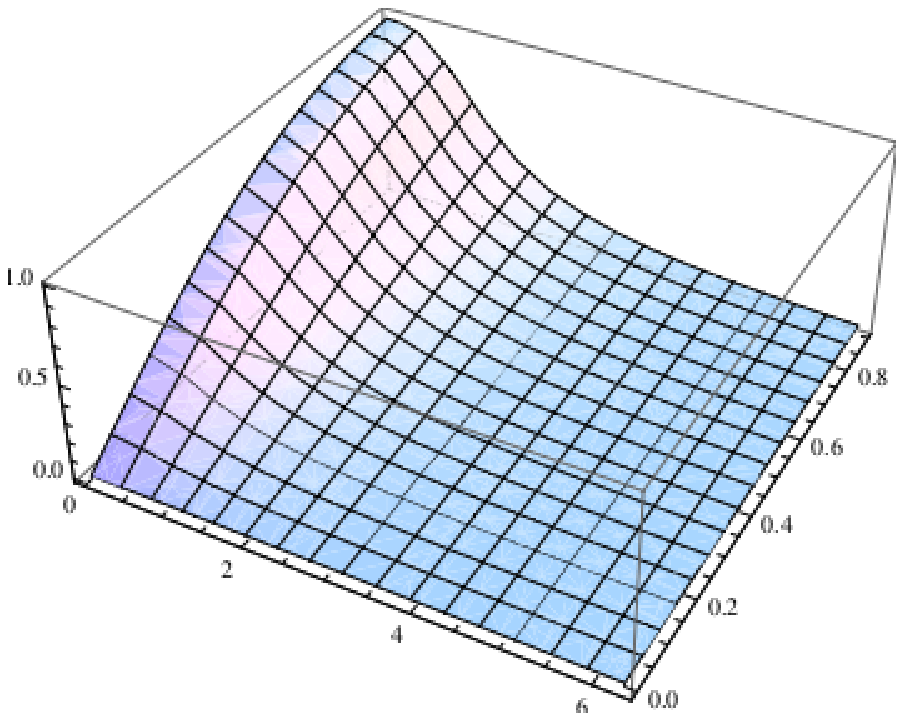}
\centerline{(a)\hskip8.1cm(b)}

\includegraphics[scale=.8]{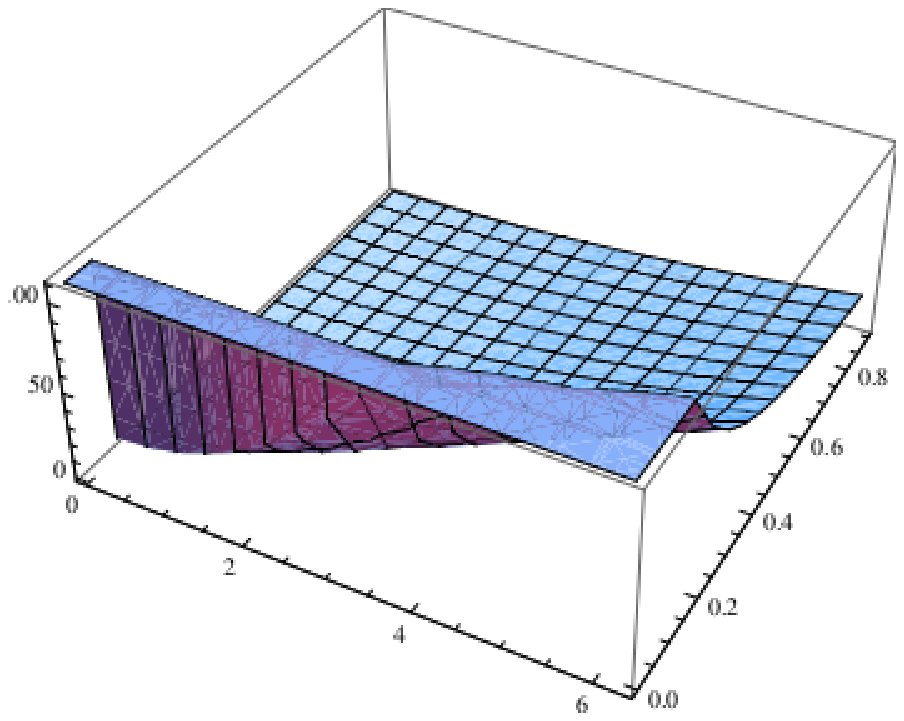}\hskip1cm\includegraphics[scale=.8]{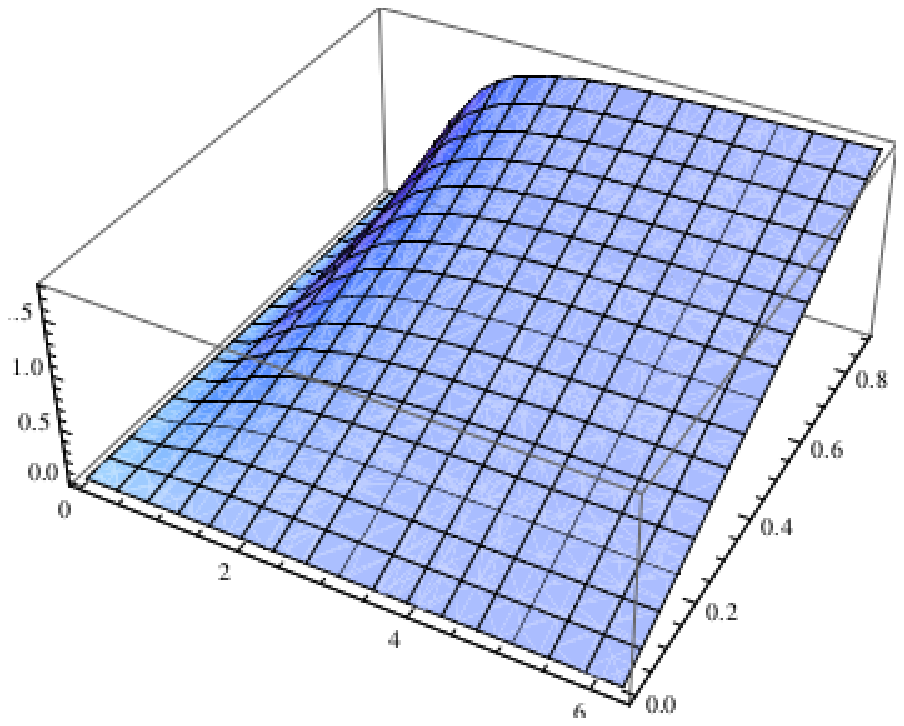}
\centerline{(c)\hskip8.1cm(d)}
\caption{The asymptotic widths, $\sigma^2_x/\ell^2$ (a), $\sigma^2_{xd}/\ell^2$ (b), $\sigma^2_p/\hbar\kappa^2$ (c) and $\sigma^2_{pd}/\hbar\kappa^2$ (d), plotted against the $(\omega_0\tau,\nu\tau)$ plane.}\label{aymptsf}
\end{figure}

One encounters a singularity at $\omega_0=0$ when the asymptotic condition $\dot\rho=0$, a set of non-linear equations for $\sigma^2_x$, $\sigma^2_d$ and $\sigma^2_{xd}$, is solved. The divergence, appearing in the first line in eqs. \eq{sigmas}, indicates that the spread of the state in the coordinate space diverges in the absence of an external potential but is kept finite by a harmonic oscillator potential. This is clearly visible in Fig. \ref{aymptsf} (a). The amount of the asymptotic drift of the particle is the result of the balance between two dissipative processes, the friction and the decoherence. The former tends to decrease the displacement while the latter increases the mobility of the particle. The equilibrium between these processes must be reached at a length scale which diverges in the free particle limit where $\sigma_x^2=\ord{t}$. The zero point fluctuations in the pure, coherent ground state of a harmonic oscillator leads to  $\sigma_x^2=\ord{\omega_0^{-1}}$, a divergence in the limit $\omega_0\to0$ which is weaker than that of the first equation in \eq{sigmas}.

An interesting impact of the singularity at $\omega_0=0$ upon the expectation value of the energy, $E=\sigma^2_p/2m+m\omega_0^2\sigma^2_x/2$, is the following. By using the results of eqs. \eq{sigmas}-\eq{sigmasp} we have
\bea
E&=&\frac{\hbar^2}{2m}\frac{d_0+d_2(\nu^2+\omega_0^2)}{2\hbar\nu}+\frac{m\omega_0^2}2\frac{\hbar(d_0+d_2\omega_0^2)}{2m^2\nu\omega_0^2}\nn
&=&\frac{\hbar}{4m\nu}[2d_0+d_2(\nu^2+2\omega_0^2)].
\eea
The $\ord{\omega_0^{-2}}$ singularity of $\sigma^2_x$, together with the $\ord{\omega_0^2}$ prefactor of the potential energy produce an $\omega_0$-independent potential energy in the first line. The non-vanishing of the potential energy in the limit $\omega_0\to0$ is a rather surprising result and is responsible of the half of the $\ord{d_0}$ contribution of the second line. Another interesting effect of the behavior  $\sigma_x^2=\ord{\omega_0^{-2}}$ is that the particle can deeply penetrate into the harmonic oscillator potential. Such a decoherence driven enhancement of the tunneling becomes a natural phenomenon by recalling that the very existence of the decoherence indicates that the particle interacts with its environment and therefore its energy is non-conserved. 

\begin{figure}
\includegraphics[scale=0.8]{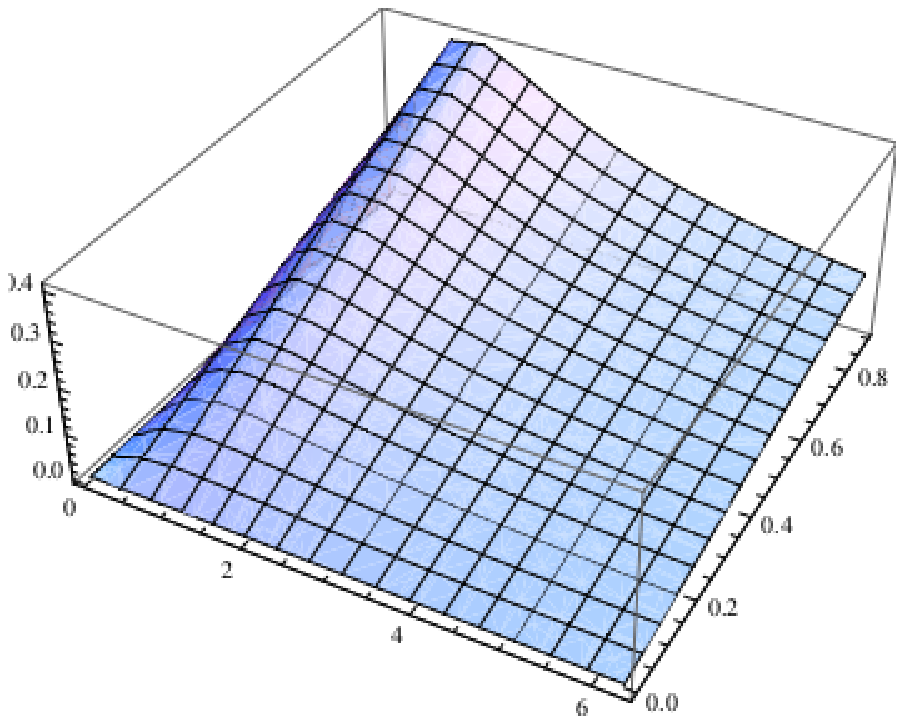}\hskip1cm\includegraphics[scale=0.8]{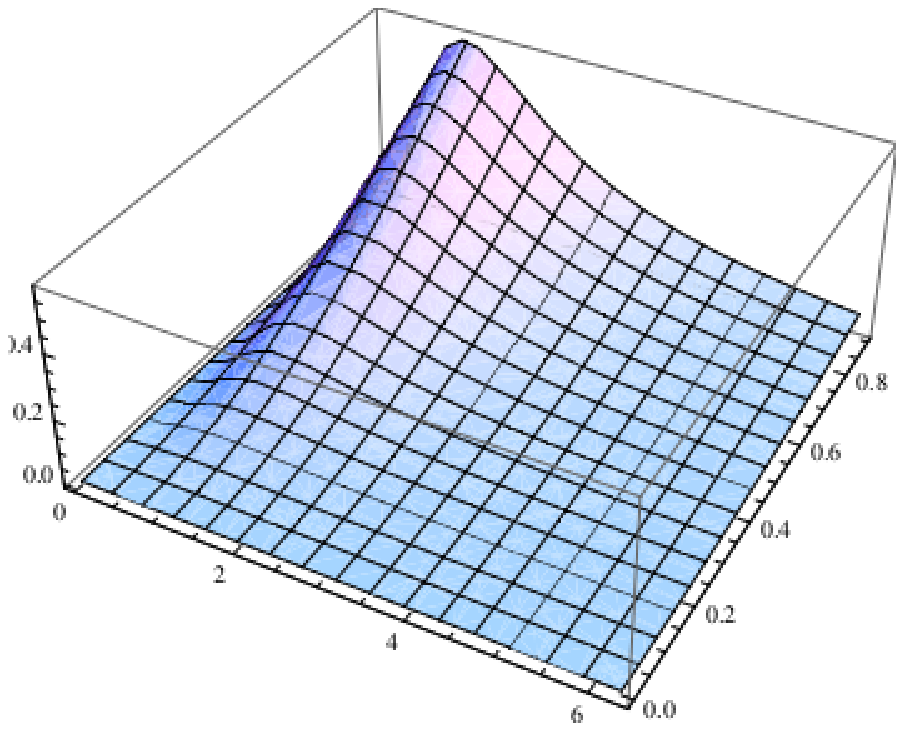}
\centerline{(a)\hskip8.1cm(b)}

\caption{The purity, (a), and the inverse uncertainty,$\hbar^2/\sigma^2_x\sigma^2_p$, (b), of the asymptotic state, plotted against the $(\omega_0\tau,\nu\tau)$ plane. Both quantities assume their maximum as the function of $\omega_0$ at $\omega^2_{max}=\sqrt{d_0(d_0+d_2\nu)}/d_2$.}\label{uncertasf}
\end{figure}

The purity of the asymptotic state, shown on Fig. \ref{uncertasf} (a), is small if the coordinate or the momentum is strongly decohered. The maximum, as the function of $\omega_0$ separates two regimes: the $\omega_0$-dependence is suppressed by $\sigma_x$ for $d_2\omega_0^2\ll d_0$ and by $\sigma_{xd}$ when $d_2\omega_0^2\gg d_0$.  The inverse of the uncertainty $\sigma^2_x\sigma^2_p$, shown in Fig. \ref{uncertasf} (b), displays a similar structure: the uncertainty is large for strongly mixed state. Table \ref{parsumt} summarizes the qualitative features of the asymptotic state manifold.

\begin{table}
\caption{The qualitative features of the asymptotic state manifold. The tilted arrows indicate the increase or the decrease as the function of the friction force or the harmonic frequency, the horizontal arrow indicates saturation.}\label{parsumt}
\begin{ruledtabular}
\begin{tabular}{lcccccr}
&Localization in $x$&Localization in $p$&Decoherence of $x$&Decoherence of $p$&Uncertainty&Purity\\
\hline
$\nu$&$\nearrow$&$\nearrow$&$\searrow$&$\searrow$&$\searrow$&$\nearrow$\\
$\omega_0$&$\nearrow$&$\searrow$&$\nearrow$&$\searrow_\to$&$\searrow\nearrow$&$\nearrow\searrow$\\
\end{tabular}
\end{ruledtabular}
\end{table}

It has been noted above that the friction increases the localization simultaneously in coordinate and momentum space. This feature must change at stronger friction to avoid the violation of the lower bound, $\sigma^2_x\sigma^2_p\ge\hbar^2$. The way this happens can simpler be seen by inspecting the localization in space and inquiring whether $\sigma^2_x$ may decrease below its value in the ground state of the harmonic oscillator, a limit which is assured by the uncertainty principle. Let us consider for this end the ratio
\be\label{assigmax}
\xi=\frac{\sigma_x^2}{\ell^2_{HO}}=\frac{d_0+\omega_0^2d_2}{m\nu\omega_0},
\ee
with $\sigma^2_{HO}=\hbar/2m\omega_0$ which indeed has a lower bound,
\be\label{lowb}
\xi_{min}=\frac{d_0+\omega_0^2d_2}{2\omega_0}\sqrt{\frac{m+4d_2}{md_0d_2}}
\ee
(c.f. the inequality \eq{positin}). $\xi_{min}$ is an increasing function of $g$ for a test particle in a gas and its minimum is reached in the limit $g\to0$ where the parametrization $d_0=d\cos\alpha$, $d_2=(d/\omega_0^2)\sin\alpha$ yields
\be
\xi_{min}=\frac{\cos\alpha+\sin\alpha}{2\sqrt{\cos\alpha\sin\alpha}}\ge1,
\ee
the lower limit being reached at $\alpha=\pi/4$. The minimum width, given by the ground state, is respected by making up the further reduction of $\sigma_x^2$, predicted by the first line of \eq{sigmas}, in a physically unacceptable manner, due to the use of negative probabilities. Note that an arbitrarily weak interaction with the environment generates a finite modification of the asymptotic state.

\section{Conclusion}\label{concls}
The time evolution of a Gaussian wave packet of a particle, moving in a harmonic potential and being subject of a friction force and decoherence, was studied in this work. Though the expectation values of the coordinate and the momentum follow the classical trajectory the second moments display non-trivial quantum fluctuations. The density matrix relaxes to an asymptotic state which is attractive in the space of Gaussian initial density matrices. The friction increases the localization and decreases the decoherence for both the coordinate and the momentum. The dependence on the oscillator frequency is more involved. The second moments respond in the coordinate and the momentum space in the opposite manner: the oscillator potential strengthen the localization and the decoherence for the coordinate but spreads and recohers the state in the momentum space. The strongly localized or delocalized initial states of the harmonic oscillator lead to qualitatively similar time dependence when shifted in time by a half period length. The final density matrix of a free particle is fully decohered in the momentum basis and is given by a simple Gibbs operator, corresponding temperature of quantum origin. The product of the uncertainties in the coordinate and the momentum spaces is a non-monotonic function of the oscillator frequency; the minimal uncertainty is reached by the maximal purity states. Finally, the purity is increasing with the amount of friction.

The asymptotic density matrix displays singularities. It changes in a discontinuous manner when a parameter of the Lagrangian is sent to zero. In particular the one of the harmonic oscillator does not converge to that of the free particle in the limit of zero oscillator frequency. Another singularity is revealed when the asymptotic state is considered as a function of the original system-environment interaction strength: the parameters of the density matrix converge to  nontrivial values even if the system-environment coupling constant tends to zero. 

Irreversibility is encountered on two different levels, it is encoded in the real and the imaginary part of the effective Lagrangian, leading to friction forces and decoherence, respectively.

These results raise several questions of which we mention but a few. How can friction weaken decoherence and make the asymptotic state even more pure?  Why does the harmonic potential contribute to decoherence in the coordinate space and to recoherence the momentum representation? Can we find freely moving particles under friction force and decoherence in Nature in the light of the result that an arbitrary weak harmonic potential modifies the asymptotic state by a finite amount? 

There are more fundamental problems, related to the way the effective dynamics of a freely moving test particle in a gas is derived by using the Landau-Ginzburg double expansion. One problem is that the parameters of such an effective theory contain the environment temperature in a rather hidden manner, for instance in the form of a loop integral whose integrand contains, among other factors, a finite temperature propagator \cite{gas}. Therefore it seems to be natural that the relaxed, asymptotic Gaussian density matrix does not reflect the environment temperature in an explicit manner. How can one recover thermalization ie. a common temperature with the environment within an effective theory, obtained via the Landau-Ginsburg expansion scheme? 

Another issue, raised by the Landau-Ginzburg expansion, is that the parameters of the effective Lagrangian arising from the underlying  microscopic dynamics determine the order of magnitude of the time scales of the friction and the decoherence. The simple dimensional argument, based on the microscopic parameters of the Lagrangian and extended to macroscopic scales may produce strongly separated relaxation and dissipation scales \cite{zurek,joos}. But one should bear in mind that both the influence Lagrangian and the master equation correspond to microscopic scales and one has to retain the higher order terms in $x^d$ to find the scales of the macroscopic regime.  Yet another open question, not considered here, is the possible existence of other, non-Gaussian asymptotic states.

One finds some hints among the results, pointing towards some well known, open problems. It is believed that a weak system-environment interaction is sufficient to generate the classical limit. The finding that an infinitesimal system-environment interaction is sufficient to leave finite trace on the asymptotic state seems to support this view. Another issue concerns the description of instabilities in many-body system, an interesting and challenging chapter of quantum field theory. If the coordinate of the harmonic oscillator is identified with a Fourier component of a quantum field then the discontinuous limit of vanishing friction force draws the attention to the importance of retaining the dissipative forces beyond the usual treatment based on the finite life-time given by the complex self energy.

\acknowledgments
I thank J\'anos Hajdu for encouragement and several discussions.

\end{document}